# Cygnus A and Other 3CR Sources in Cosmological Tests

**By Erick J. Guerra & Ruth A. Daly**

Joseph Henry Laboratories, Princeton University, Princeton, NJ, USA

Powerful extended radio galaxies like Cygnus A can be used for cosmological tests. The characteristic, time-independent length $D_*$ for several radio sources is calculated and compared to the average physical length for a sample of radio galaxies. The ratio of these two lengths as a function of redshift is used to constrain cosmological parameters. Cygnus A is the only low- redshift ($z < 0.5$) radio galaxy for which we have an estimate for $D_*$. We comment on the sensitivity to this one low-redshift point, and results including and excluding Cygnus A are shown. A low density universe is favored, and the one free model parameter is relatively well constrained.

## 1. Introduction

The use of the physical sizes of radio sources as a cosmological test has been considered by many [1,2,3,4] since radio sources are observed to fairly high redshift. The coordinate distance $(a_o r)$, used to calculate the size of a source is a function of the density parameter $\Omega_o = 8\pi G\rho_o/3{H_o}^2$, and the normalized cosmological constant $\lambda_o = \Lambda/3{H_o}^2$, where $\rho_o$ is the mean mass density of the universe, $H_o$ is the Hubble constant, and $\Lambda$ is the cosmological constant. It has been shown that radio sources are not, strictly speaking, standard yardsticks [4]. However, if the evolution with redshift of physical sizes could be predicted and compared to observations, then radio sources could be used for cosmography [5].

Daly [6] presents a method where a characteristic size $D_*$ (called $l_*$ by Daly [6,7]) for a powerful extended radio source can be calculated from observables assuming a value for one model parameter $\beta$. The evolution of these chracteristic sizes for a chosen sample should, for the correct parameters ($\beta$, $\Omega_o$, $\lambda_o$), track the evolution of the average physical size of a large sample of similar sources (see §2). The cosmological dependence of the ratio of these sizes goes as a power greater than one of $(a_o r)$ for realistic values of $\beta$ (see §4). Thus, as we compare samples at high redshift, the dependence of $(a_o r)$ on $\Omega_o$ and $\lambda_o$ will allow us to constrain the cosmological parameters. It should be noted that the results of this analysis are insensitive to the value of $H_o$ (see §4); here $H_o$ is parameterized in the usual way: $H_o = 100\,h$ km s$^{-1}$ Mpc$^{-1}$.

To calculate $D_*$ for a radio source, one must have the lobe propogation velocity $v_L$, the lobe magnetic field strength $B$, and the lobe width $a_L$ (see §2). These parameters can be extracted from multi-frequency observations of radio sources [8,9].

## 2. The Characteristic Size $D_*$ and the True Physical Size $D$

Daly [6] describes how to calculate a characteristic, time-independent length scale for a radio source. If a radio source has constant lobe propogation velocity $v_L$ with a stable direction of outflow, and is active for a lifetime $t_*$, the characteristic length is

$$D_* = v_L t_*. \tag{2.1}$$

One can think of $D_*$ as the core-lobe size at the end of the source's lifetime, or as the average lobe-lobe separation. This method may be applied to sources that are supersonic propagators with no significant backflow or bridge distortions, such as the high power FRII radio galaxies (see §3 and [6]). Note that present observations are consistent with $v_L$ being roughly constant over the lifetime of a particular source [7,9,10].

If the luminosity in directed kinetic energy, $L_j$, is roughly constant over the lifetime of the source, then $t_* \simeq E_i/L_j$ where $E_i$ is the initial energy of the central engine that powers the jets. Note that $E_i$ and $L_j$ will depend only on processes occuring near or at the central engine. A power law is assumed between $t_*$ and $L_j$

$$t_* \propto {L_j}^{-\beta/3}, \tag{2.2}$$

which is equivalent to assuming $E_i \propto {L_j}^{1-\beta/3}$ (see [6]). The determination of $\beta$ has implications for models of energy extraction from the central engine [6,7,11]. Combining eqs. (2.1) and (2.2), it is clear that

$$D_* \propto v_L {L_j}^{-\beta/3}. \tag{2.3}$$

The radio lobe propagates supersonically and does $P\,dV$ work on the ambient medium [12]. The pressure $P$ of the lobe is given by the strong shock jump conditions $P = \frac{3}{4}\rho_a {v_L}^2$ where $\rho_a$ is the mass density of the ambient medium. The energy input to the hotspot and lobe in a time interval $\delta t$ is $\delta E \simeq L_j\,\delta t$. Equating $\delta E$ to $P\,dV$ with volume element $dV = \pi {a_L}^2 v_L\,\delta t$ gives a relation between $v_L$ and $L_j$:

$$L_j \propto n_a {a_L}^2 {v_L}^3 \propto \left(\frac{v_L}{k}\right)^3 \tag{2.4}$$

where $n_a$ is the number density of the ambient media and

$$k \equiv \left(n_a {a_L}^2\right)^{-1/3}. \tag{2.5}$$

Combining eqs. (2.3) and (2.4),

$$D_* \propto k^\beta {v_L}^{1-\beta}. \tag{2.6}$$

For supersonic flow, the lobe is ram pressure confined [13], so the pressure of the lobe $P \propto n_a {v_L}^2$. The lobe pressure will be $\propto B^2$ if the magnetic field $B$ and the relativistic electrons are near equipartition. Thus, the ambient gas density

can be estimated as [6]

$$n_a \propto (B/v_L)^2. \tag{2.7}$$

Combining eqs. (2.5),(2.6), and (2.7), gives an expression for $D_*$ in terms of observables.

$$D_* \propto \left(\frac{v_L}{B a_L}\right)^{2\beta/3} {v_L}^{1-\beta}. \tag{2.8}$$

For our purposes, the normalization of $D_*$ is unimportant and arbitrary. Note that in theory, $D_*$ is time-independent for a given source and thus is the same no matter when the source is observed during its lifetime.

The core-lobe size of a radio source $D$ is given by $D = v_L t$ where $t$ is the age of the source and $v_L$ is assumed to be constant. Thus for a given source $D = (t/t_*)D_*$. For the sources we consider, the lobe-lobe size is well approximated by $2D$ (see §4).

In a given redshift range, one expects the $D_*$'s for a sample to be peaked around some central value. Taking the $D_*$'s in this redshift range to be approximately equal to a constant, $D_z$, it follows that

$$\overline{D} = \overline{\left(\frac{t}{t_*}\right) D_*} \simeq \overline{\left(\frac{t}{t_*}\right)}\, D_z. \tag{2.9}$$

If we assume the distribution of $t/t_*$, the fraction of a source's lifetime at which it is observed, is just a constant for a given sample, then $\overline{D} \simeq D_z/2$. More generally, if the $t/t_*$ distribution of a sample does not evolve with redshift, then $\overline{D} \propto D_z$.

## 3. Samples

In order to use powerful extended radio sources for cosmology, the comparison sample must be as homogenous as possible. Thus, only radio galaxies with $178 MHz$ power $P_{178}(q_o = 0) \gtrsim 3h^{-2} \times 10^{26}$ W Hz$^{-1}$ sr$^{-1}$ are considered [11]. The power cut insures that backflows and bridge distortions are not significant, and that the lobes are supersonic [6,10,14]. Although this method might be extended to radio-loud quasars, current observations favor the use of radio galaxies [6].

The radio galaxies used for the comparison sample are the 3CR radio galaxies with FRII morphology listed in Table 1 of McCarthy *et al.* [15], subject to the power constraint mentioned above. To calculate the radio power of these sources, fluxes and spectral indices from Spinrad *et al.* [16] are used. Cygnus A, which was not listed by McCarthy *et al.* due to its low galactic latitude, is included in this study since it is the closest high-power FRII radio galaxy, and is well-studied.

To compute $D_*$, values for $v_L$, $a_L$, and $B$ are needed, and those extracted by Wellman & Daly [17,18] are used here. Wellman & Daly reanalyzed the published maps of 14 radio galaxies, from [8,9], to obtain consistent values for the parameters needed in eq. (2.8). These sources are only a subset of the comparison sample, and in the future, $D_*$'s for more of this sample should be obtained [11].

| $z$ range | # sources | (a) | (b) | (c) | (d) | (e) | (f) |
|---|---|---|---|---|---|---|---|
| | | ($kpc/h$) | ($kpc/h$) | ($kpc/h$) | ($kpc/h$) | ($kpc/h$) | ($kpc/h$) |
| 0.0-0.3 | 3 | 33±5 | 66±14 | 34±4 | 68±14 | 36±4 | 72±13 |
| 0.3-0.6 | 15 | 89±15 | 177±42 | 98±17 | 197±47 | 114±19 | 227±54 |
| 0.6-0.9 | 27 | 63±7 | 126±18 | 74±8 | 149±21 | 90±9 | 179±25 |
| 0.9-1.2 | 19 | 46±8 | 91±21 | 57±10 | 114±27 | 70±12 | 141±33 |
| 1.2-1.6 | 10 | 41±11 | 83±30 | 55±14 | 110±40 | 69±17 | 137±50 |
| 1.6-2.0 | 8 | 25±6 | 50±17 | 36±8 | 71±24 | 44±10 | 88 ±30 |

Table 1. Average physical sizes by redshift bin for the sample described in §3: (a) Core-Lobe and (b) Lobe-Lobe for ($\Omega$,$\lambda_o$)=(1.0,0); (c) Core-Lobe and (d) Lobe-Lobe for ($\Omega$,$\lambda_o$)=(0.1,0); (e) Core-Lobe and (f) Lobe-Lobe for ($\Omega$,$\lambda_o$)=(0.1,0.9).

## 4. Calculations & Results

Redshift bins were chosen to be of similar size, and the average core-lobe and lobe-lobe sizes for the whole comparison sample are listed in Table 1 for ($\Omega_o$,$\lambda_o$)=(1.0,0), (0.1,0), & (0.1,0.9). The average lobe-lobe sizes are about twice the average core-lobe sizes, as expected. Figure 1 shows the average core-lobe size for the sample. Except for the lowest redshift bin, the average size of these radio galaxies decreases with increasing redshift.

Each redshift bin has an average physical size $\overline{D}$. The data needed to calculate $D_*$ are only available for small number of sources, and the determination of typical $D_*$ for a bin based on one or two sources could be misleading [11]. In order to compare the evolution of $D_*$ to $\overline{D}$, the ratio $\overline{D}/D_*$ is calculated for each lobe and then each source, using $\overline{D}$ from the appropiate redshift bin.

In principle, for the correct values of $\beta$ and the cosmological parameters, the ratio $\overline{D}/D_*$ is constant and independent of redshift. For example, the ratio $\overline{D}/D_*$ assuming $\beta = 2.0$ and ($\Omega_o$,$\lambda_o$)=(0.1,0) is shown in Figure 2 for each lobe with a calculated $D_*$. The ratio in Figure 2 is nearly constant for all redshift, and similar results are found for different cosmological parameters [11].

At a given redshift, $\overline{D}/D_* \propto (a_o r)^{3/7+2\beta/3}$ [6], and for $\beta > 1$, the ratio should depend on $(a_o r)$ to a power greater than one. Since the normalization of the ratio is arbitrary, and $(a_o r) = {H_o}^{-1} f(\Omega_o, \Lambda, z)$, $H_o$ does not enter in the analysis.

Using the three sets of cosmological parameters mentioned above, and letting $\beta$ vary, $\overline{D}/D_*$ is fitted to a constant using the weighted mean, and a reduced $\chi^2$ is calculated. The values for the reduced $\chi^2$ as a function of $\beta$ including the Cygnus A point are shown in Figures 3 & 4 for the core-lobe and lobe-lobe calculations respectively.

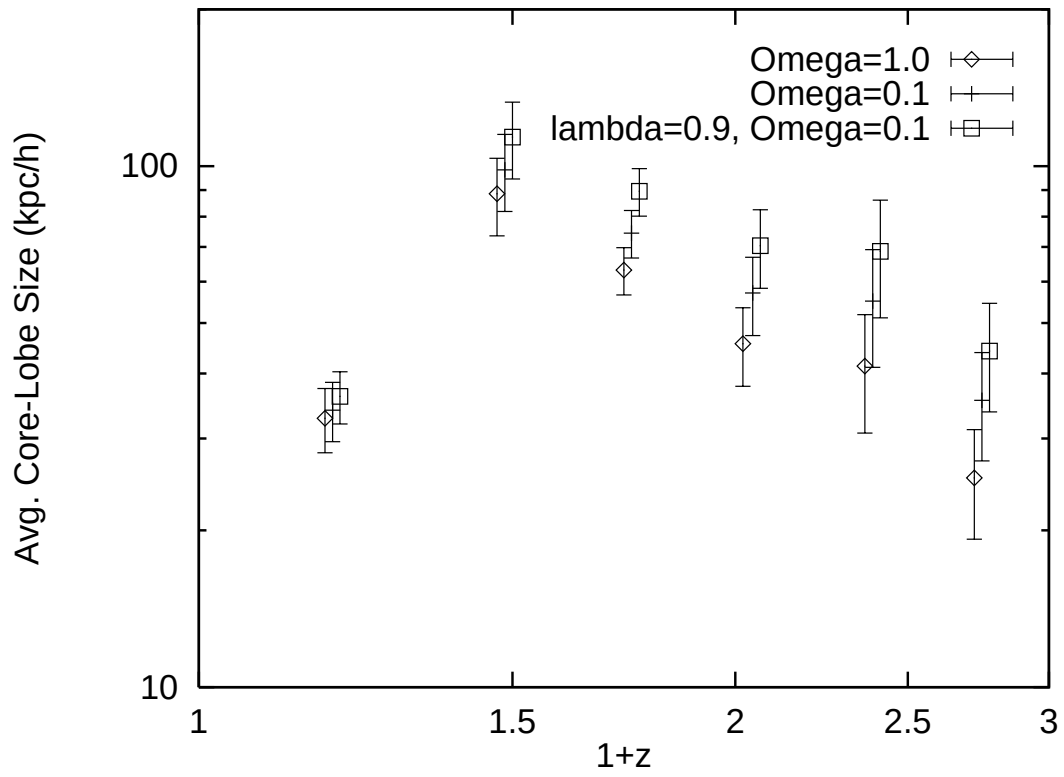

FIGURE 1. The average core-lobe size, $\overline{D}$, for the sample in Table 1.

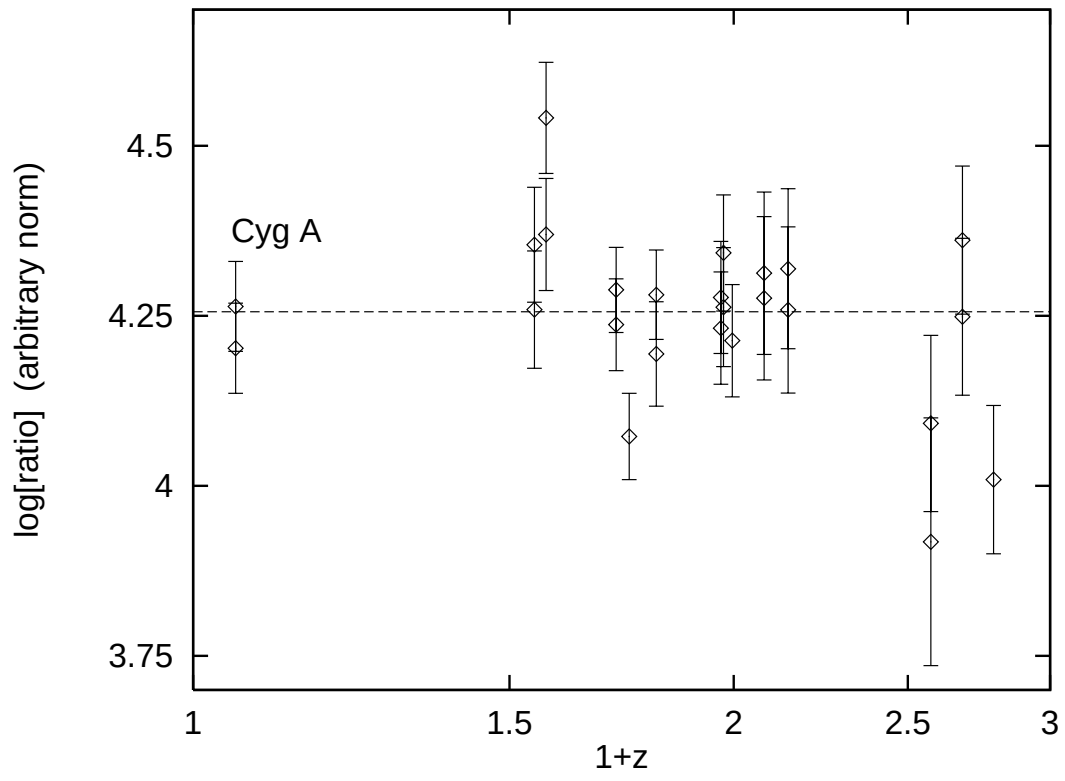

FIGURE 2. $\overline{D}/D_*$ for each lobe with $\beta = 2.0$ and $(\Omega_o,\lambda_o)$=(0.1,0).

The best-fit value of $\beta$ is about two in each of the three cosmologies studied. This is consistent with the result obtained by Daly [6,7], $\beta = 1.5 \pm 0.5$, based on a different argument. In calculating the reduced $\chi^2$ for the core-lobe comparison, the two lobes of a given source are treated as statistically independent which is not, strictly speaking, correct, whereas the lobe-lobe comparison is valid statistically [11].

A low $\Omega_o$ universe is favored over a flat, matter-dominated universe as can be seen in both Figures 3 & 4. In principle, this method is sensitive to the difference between a $\Lambda$-dominated or curvature-dominated cosmology, but both of these models fit the current data well [11].

Figures 5 & 6 are to be compared to Figures 3 & 4, and show the reduced $\chi^2$'s excluding Cygnus A. This eliminates the lowest redshift bin, which has the fewest sources to calculate $\overline{D}$. These figures show that the results are insensitive

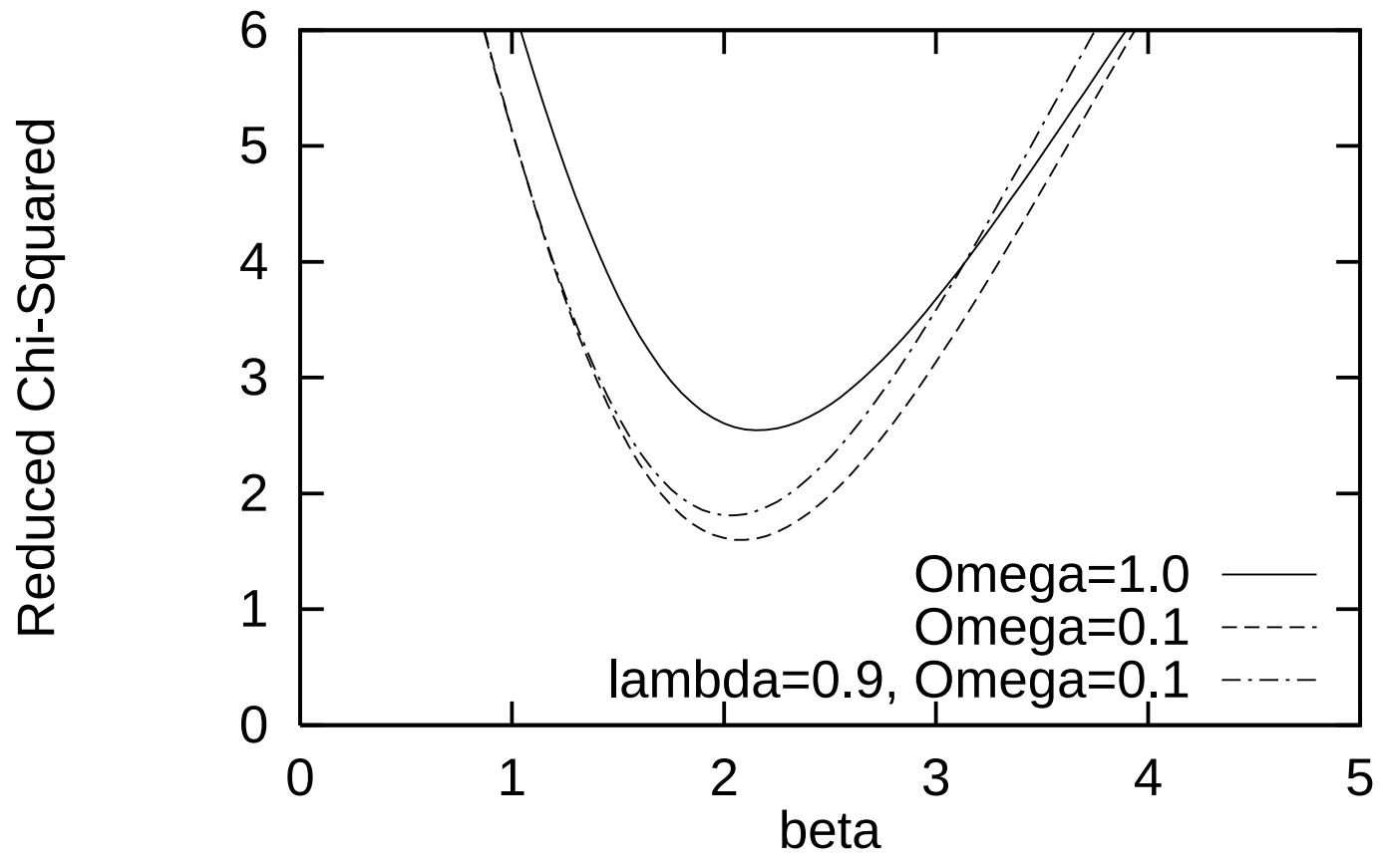

FIGURE 3. Reduced Chi-squared for Core-Lobe comparison including Cygnus A.

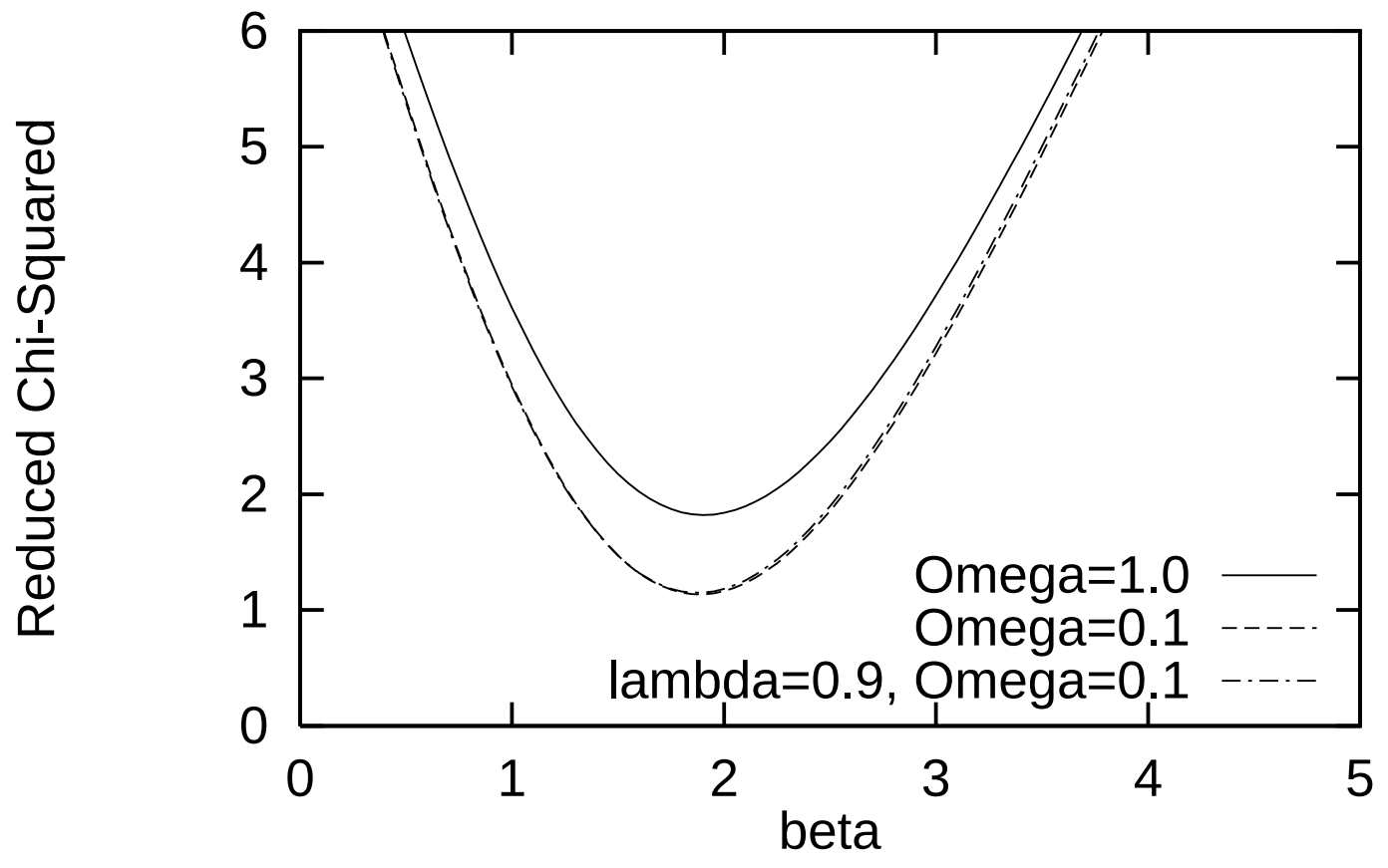

FIGURE 4. Reduced Chi-squared for Lobe-Lobe comparison including Cygnus A.

to the lowest redshift bin. $\overline{D}$ for the lowest redshift bin and $D_*$ for Cygnus A are both low compared to the adjacent redshift bin, yet the ratios are comparable to those at higher redshift for $\beta \simeq 2$ [11], as is evident in Figure 2.

More details of this analysis and additional figures are given by [11].

## 5. Conclusions

The results presented here indicate the promise of this method. The fact that Cygnus A seems to fit in the sample in terms of this analysis and other properties, such as $L_j$ and $n_a$ [6,7] is encouraging. This analysis favors a low-density universe that is curvature-dominated or $\Lambda$-dominated over a flat, matter-dominated universe.

Computing $D_*$'s for more sources and increasing the size of the comparison sample are essential for the use of this method to estimate cosmological parameters and to test the model for internal consistency. Thus, a long term goal is to expand the number and redshift range of both samples.

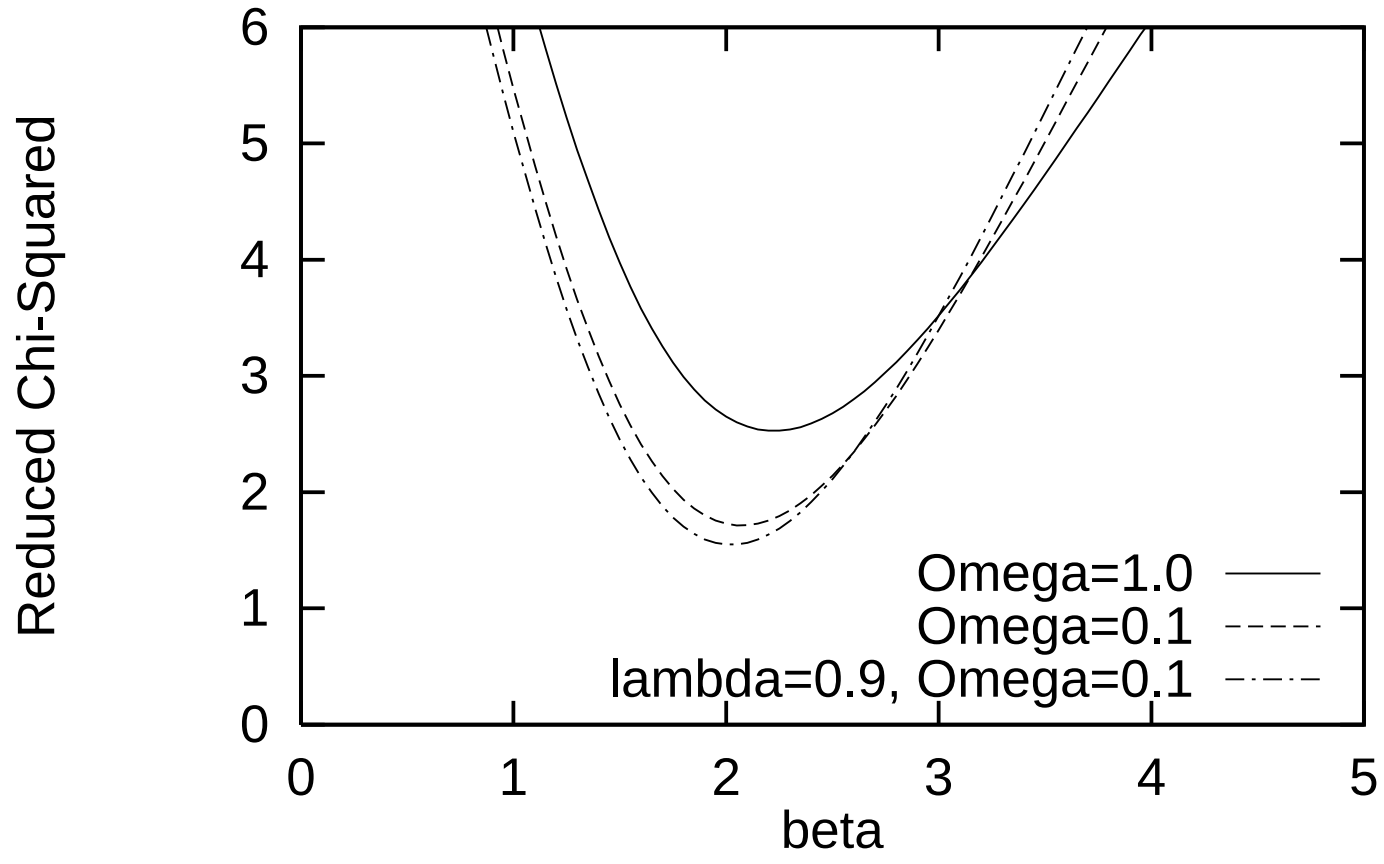

FIGURE 5. Reduced Chi-squared for Core-Lobe comparison excluding Cygnus A.

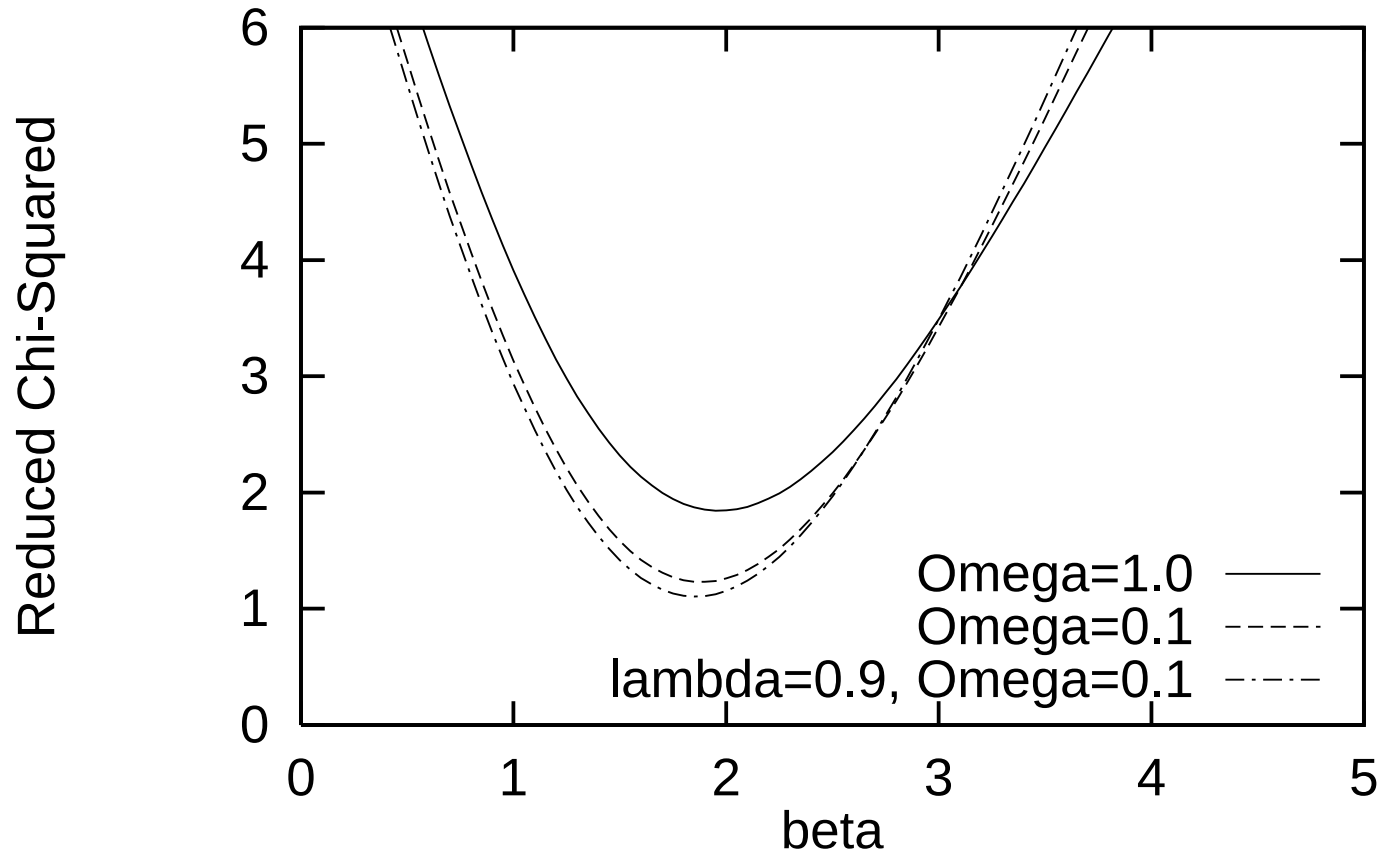

FIGURE 6. Reduced Chi-squared for Lobe-Lobe comparison excluding Cygnus A.

The authors would like to thank Lyman Page and Lin Wan for helpful discussions. A special thanks goes to Greg Wellman for access to his data and for valuable discussions. This work was supported by an NSF Graduate Fellowship, an NSF National Young Investigator Award, and the U.S. National Science Foundation.